\documentclass[runningheads]{llncs}
\usepackage{graphicx}
\usepackage{mathtools}
\usepackage[dvipsnames]{xcolor}
\usepackage{amsmath}
\usepackage{amssymb}
\usepackage{hyperref}
\begin{document}
\title{Scalable Modeling of Nonlinear Network Dynamics in Neurodegenerative Disease}
\titlerunning{COMIND}

% \author{Daniel Semchin\inst{1} \and Emile d'Angremont\inst{2} \and Marco Lorenzi\inst{3} \and Boris Gutman\inst{1} }

% \institute{$^{1}$Illinois Institute of Technology\samelineand
% $ ^{2}$Amsterdam UMC, Vrije Universiteit Amsterdam\samelineand
% $ ^{3}$Universite of Cote D'Azur}

\author{Daniel Semchin\inst{1}  \and
Emile d'Angremont\inst{2} \and Marco Lorenzi\inst{3} \and
Boris A. Gutman \inst{1}}
\authorrunning{D. Semchin E. d'Angremont M. Lorenzi B. Gutman}
\institute{Illinois Institute of Technology, Department of Biomedical Engineering, USA \and
Amsterdam University Medical Center, Department of Anatomy and Neurosciences, The Netherlands \and
Epione Team, Inria Center of Universit\'e C\^ote d’Azur, Sophia Antipolis, France
\\
\email{dsemchin@hawk.illinoistech.edu}}

\maketitle           
\begin{abstract}

Mechanistic models of progressive neurodegeneration offer great potential utility for clinical use and novel treatment development. Toward this end, several connectome-informed models of neuroimaging biomarkers have been proposed. However, these models typically do not scale well beyond a small number of biomarkers due to heterogeneity in individual disease trajectories and a large number of parameters. To address this, we introduce the Connectome-based Monotonic Inference of Neurodegenerative Dynamics (COMIND). The model combines concepts from diffusion and logistic models with structural brain connectivity. This guarantees monotonic disease trajectories while maintaining a limited number of parameters to improve scalability. We evaluate our model on simulated data as well as on the Parkinson's Progressive Markers Initiative (PPMI) data. Our model generalizes to anatomical imaging representations from a standard brain atlas without the need to reduce biomarker number.

%%%%%%%%% TODO %%%%%%%%%%%%%%%%%%%  
\keywords{Disease Progression Modeling \and Brain Connectivity \and Neurodegenerative Disease}
\end{abstract}

\section{Introduction}

Recent advances in data-driven disease progression modeling (DPM) have significantly enhanced our understanding of neurodegenerative diseases. Such models promise broad applicability, both as simulators of treatment effect and to estimate pre-symptomatic time of disease onset. DPMs and related inference algorithms, i.e. methods inferring long-term biomarker trajectories from observations over a short time period, can be broadly classed along several dichotomies, e.g.: discrete- vs. continuous-time, mechanistic vs. purely generative statistical, monotonic vs. non-monotonic, subtyped vs. single-mode. Other ways to classify DPM models probably exist, but we find these four most useful for contextualizing this work. 
Discrete-time methods are largely derivates of the Event-Based Model (EBM) \cite{1-EBM}. EBM assumes monotonic biomarker trajectories, described by an ordering of discrete stages. The method works well in cross-sectional datasets. A widely used subtyping derivative of EBM, Subtype and Stage Inference (SuStaIn) \cite{2-SusTain} seeks to disentangle temporal and phenotypic heterogeneity to identify population subgroups with common patterns of disease progression. The method groups individuals with common canonical biomarker orderings, defining disease stages independently for each subtype. A continuous-time alternative, the DIVE model \cite{3_DIVE} infers coherent spatiotemporal patterns using a parametric model of progression. DIVE both segments the cortex and finds cluster-level parameters of logistic progression curves, requiring longitudinal data for stable inference. Disease Progression Modelling and Stratification (DP-MoSt) \cite{4_DP_Most} also fits logistic curves, simultaneously subtyping biomarker trajectories and samples into discrete patient groups. More recently, approaches like Brain Latent Progression (BrLP) \cite{5_BrLP} have integrated latent diffusion models with prior knowledge to enhance spatiotemporal disease progression predictions. DIVE and DP-MoSt both assume monotonic biomarker change. With the exception of DIVE, these approaches primarily operate on predefined regions of interest and no method above explicitly incorporates the brain's connectivity structure, limiting their ability to capture mechanistic aspects of disease spread. 
Complementing the subtyping approaches, network-based models have emerged as powerful tools for understanding disease progression through the lens of brain connectivity. These approaches are supported by neuropathological evidence for "prion-like" transsynaptic transmission of disease agents like misfolded tau and beta amyloid, suggesting that disease is transmitted along neuronal pathways rather than by proximity. An early example, the Network Diffusion Model (NDM) \cite{6_NDM} models disease transmission as a diffusive mechanism over a tractography-based connectome. NDM demonstrated significantly higher accuracy in predicting longitudinal patterns of atrophy and metabolism compared to non-network-based models. Generalizing this, the Accumulation-Clearance-Propagation (ACP) \cite{7_ACP} model expands the connectivity-based propagation idea with a three-compartment model allowing for time-varying biomarker diffusion effects that represent distinct processes. These and related  \cite{8_NODE_progression} progression models explicitly use a system of differential equations; particularly when coupled with empirically established structural connectivity, the models provide a mechanistic explanation for disease spread in the brain. However, most current implementations are limited in their scalability to the full brain connectome with dozens or hundreds of regions.

As highlighted in recent reviews, data-driven disease progression models provide unique advantages over 'black box' machine learning tools by being inherently interpretable and requiring fewer data \cite{13_Young_DPM_review,14_Moravveji_DPM_review}. However, a critical gap remains in our current modeling arsenal: the lack of approaches that simultaneously leverage the brain's connectivity structure, scale to large numbers of brain regions, accommodate disease heterogeneity, and provide mechanistic interpretability over extended time ranges. Furthermore, the scale of modern neuroimaging datasets and high-resolution connectome mapping requires models that can efficiently operate across dozens to hundreds of brain regions while maintaining computational tractability. Existing approaches either sacrifice scale for mechanistic detail or interpretability for computational efficiency, limiting their practical application.

The approach described here - the Connectome-based Monotonic Inference
of Neurodegenerative Dynamics (COMIND) - represents a step to address these requirements by developing a network-connectome based disease progression model that scales to large numbers of brain regions while providing interpretable, mechanistic insights into disease evolution over extended time periods. Our method improves scalability of network-based mechanistic modeling, offering new opportunities for understanding the complex spatiotemporal dynamics of neurodegenerative disease progression.

\section{Methods}

A number of the proposed progression models have used logistic evolution as the basic model-building element. The idea is appealing as it mirrors our intuition that neurodegeneration has a gradual onset, a period of rapid advance, and a saturation point. Indeed, the sigmoid is a solution to the logistic ODE $\dot{x}=k(1-x)x$, a dynamic in which the rate of accumulation is a product of self-supply and capacity. Our motivation is to incorporate network dynamics into this idea, while maintaining the asymptotic properties of the univariate logistic ODE. 

\subsection{Dynamic Propagation Model}

We begin by considering a dynamic $p$-dimensional system of accumulating neurodegenerative effects with a static transition matrix $K\in\mathbb{R}^{p\times p}$ defined in some way by brain connectivity. As in a standard logistic differential equation, we define the rate of neurodegenerative accumulation as a product of the present state of neurodegeneration and remaining capacity. This can be expressed as 

\begin{equation}
\label{naive_logistic}
    \frac{d\mathbf{x}}{dt} = [I-D(\mathbf{x})]K\mathbf{x},\;\; \mathbf{x}\in[0,1]^p
\end{equation}

Here, $D(\mathbf{x})$ is the diagonal matrix with entries in $\mathbf{x}$. We consider external sources of neurodegeneration as constant, again contributing to the rate of regional accumulation in proportion to regional capacity. 

\begin{equation}
\label{logistic_forcing}
    \frac{d\mathbf{x}}{dt} = [I-D(\mathbf{x})][K\mathbf{x} + \mathbf{f}],\;\; \mathbf{x}\in[0,1]^p, \;\; \mathbf{f}\in[0,\infty)^p, 
\end{equation}

where $\mathbf{f}$ represents the constant "forcing" term in the ODE. For the $k^\text{th}$ individual regional biomarker, this is equivalent to $\dot{x}^k(t)=(1-x^k)(\sum_mK_{km}x^m + f^k)$. To reduce the number of model parameters, here we consider the transition matrix simply as a scaled version of the connectome $K^*$, i.e. $K=s_tK^*$, where $s_t$ is a parameter denoting timescale. Finally, to construct canonical trajectories of neurodegeneration we must scale solutions of the dynamic system to the natural scale of imaging biomarkers. This adds a scaling vector $\mathbf{s}\in\mathbb{R}^p$ to our trajectory parameter set. Suppose $\mathbf{x}(t)$ is the solution to \eqref{logistic_forcing}. Then the predicted biomarker value at time $t$ is $\mathbf{\hat{y}}(t)= \mathbf{s}\odot \mathbf{x}(t)$, where $\odot$ is element-wise multiplication. In summary, given a canonical connectome, our biomarker trajectory $\mathbf{\hat{y}}(t) = \mathbf{\hat{y}}(t;\theta)$ is parameterized by $\theta=\{s_t,\mathbf{s},\mathbf{f}\}$, with $2p+1$ parameters.

\subsection{Modeling Subject-Specific Time-Shift}
As is common practice, we assume a Gaussian i.i.d. noise model. For the $k^{\text{th}}$ biomarker $y_{ij}^k$ measured in subject $i$ at timepoint $j$, biomarker likelihood is expressed as 

\begin{equation}
\label{single_biomarker_like}
    P(y_{ij}^k | \theta,\beta_i)=\mathcal{N}(y_{ij}^k | \hat{y}^k(t_{ij}+\beta_i;\theta),\sigma_k).
\end{equation}

Here, $\beta_i$ is the subject-specific "disease-time" or time-shift at the initial scan, and $t_{ij}$ is the time of scan at visit $j$, given that $t_{i0}=0$. We consider the possibility that certain biomarkers, such as a subset of clinical scores, can have a privileged status. Specifically, we set aside some set of biomarkers $\mathbf{c}\in\mathbb{R}^c$ - excluded from $\mathbf{y}$ - that condition individual time-shifts by a simple linear model:

\begin{equation}
\label{clinical_prior}
    P(\beta_i | \mathbf{c}_i)=\prod_{j}\mathcal{N}(\beta_i+t_{ij} | a+\langle\ \mathbf{b},\mathbf{c}_{ij}\rangle,\sigma_c),
\end{equation}

where $a$ and $\mathbf{b}$ are some unknown regression parameters. 

\subsection{Biomarker Trajectory Conditioning and Likelihood Estimation}

To represent the fact that brain function is largely self-contained and locally protected from external inputs, we assume that the number of brain regions affected by $\mathbf{f}$ is small and therefore $\mathbf{f}$ is sparse. Additionally, to prevent exploding timescales which allow all samples to be mapped to flat trajectories, we add an $\mathit{l}2$ penalty to $s_k$. Thus, our prior on the dynamic system parameters is 

\begin{equation}
\label{theta_prior}
    P(\theta) \propto \text{exp}{[-w_f|\mathbf{f}| -w_ss_t^2]},
\end{equation}

Combining the terms in \eqref{single_biomarker_like},\eqref{clinical_prior},\eqref{theta_prior}, we have the posterior for $\theta$ and $\beta$:

\begin{equation}
\label{posterior}
    P(\theta,\beta| \mathbf{y}, \mathbf{c}) = \prod_{ijk} P(y_{ij}^k | \theta,\beta_i)\times P(\theta)\times \prod_i P(\beta_i | \mathbf{c}_i)
\end{equation}

\subsection{Model Estimation}
We estimate model parameters and subject time-shifts using a generalized Expectation - Maximization approach. We recognize that other optimization strategies may improve the estimation; however, in practice the approach proved robust even for relatively low-signal data such as anatomical imaging in early-stage Parkinson's disease. 

E-step: We assume fixed $\beta$ and optimize $\theta$ by minimizing the negative log likelihood of \eqref{posterior}. For components of $\theta$ that parameterize \eqref{logistic_forcing}, we numerically integrate both the ODE and the Jacobians of the ODE with respect to $f$, $s_t$. In practice, we find that alternating between L-BFGS and Nelder-Mead optimization methods achieves the best results.

M-step: We assume a fixed biomarker trajectory and optimize $\beta$. In the case where we assume that $\mathbf{b}\neq\mathbf{0}$, this step is further split into two substeps: 1. estimating global clinical regression parameters $a, \mathbf{b}$ and dispersion $\sigma_c$ based on current $\beta$; 2. optimizing \eqref{posterior} to update $\beta$ only. The first step is a simple linear problem. The second is a smooth convex problem, solved efficiently with the L-BFGS algorithm. 

\begin{figure}
    \centering
    \includegraphics[width=0.9\linewidth]{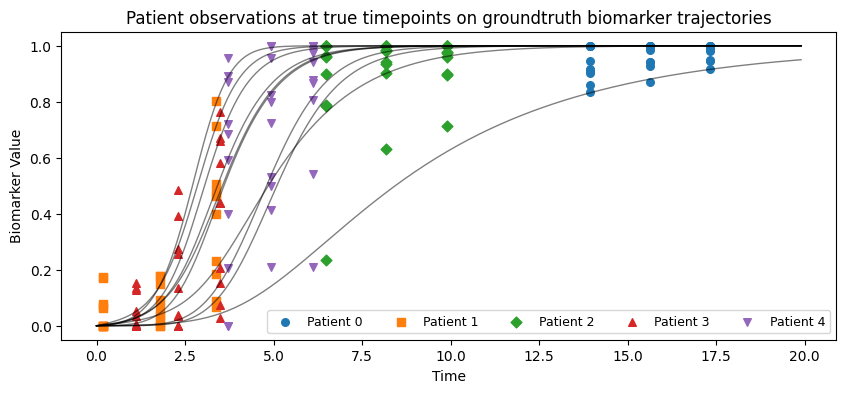}
    \caption{Five randomly generated patients with their observed biomarker values at the true time. Black lines represent true trajectories.}
    \label{fig:synthetic_scatter}
\end{figure}
\begin{figure}
    \centering
      \includegraphics[width=0.9\linewidth]{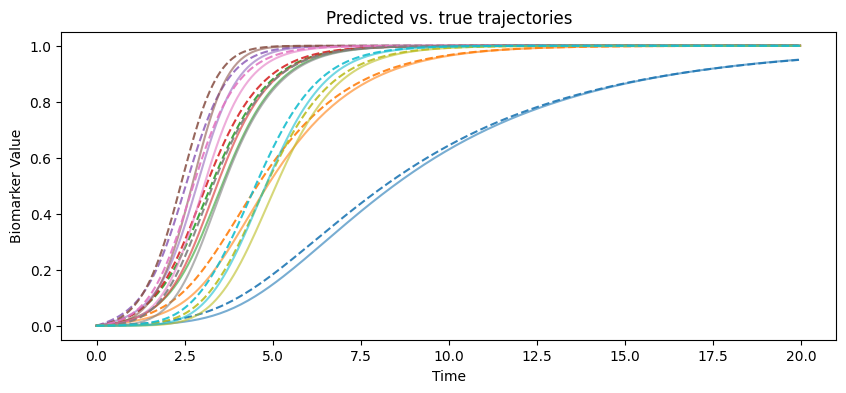}
  \caption{Fitting synthetic trajectories with the COMIND model. Solid lines are true trajectories, dashed lines are COMIND estimates.}
  \label{fig:Synthetic_trajectories}
\end{figure}

\section{Data and Experiments}

\subsubsection{Synthetic Data generation}
True disease trajectories were generated using \eqref{logistic_forcing}, with initial conditions equal to zero and $\mathbf{f}$ being a random realization drawn from Gamma(0,0.05). The connectivity matrix $K$ was generated to mimic the sparsity of the connectome, with the diagonal uniformly set to 0.

% The values of the first two off-diagonals are distributed on $\mathcal{N}(0,\,1)$  and multiplied by the reciprocal of the offset from the main diagonal. A mask is a applied to the off diagonal that randomly takes 0.1 multiplied by the offset of the diagonal value and zeros them out, meaning the further from the main diagonal the greater the sparsity. 

After generating the trajectories, 200 patients were simulated with 3 equally spaced consecutive observations per patient starting at a random time-point. The spacing for each patient is randomly distributed on Gamma(2, 0.5). We added Gaussian noise with $\sigma=0.1$ for each observed value. Each patient had an initial $\mathbf{\beta}$ drawn from a uniform distribution between 0 and $t_\text{max}$. We also generated a "privileged" clinical feature. Clinical regression parameters were randomly chosen between [1,5] for slope and [0,10] for the bias term. A value representing a single clinical score was then generated for each observation as follows: $\text{cognitive score} = a*(t_{ij} + \mathcal{N}(0,\,1)) + b$. 
Examples of trajectory and sample generation are seen in $\textbf{Fig.\ref{fig:synthetic_scatter}}$.

% \begin{figure}
%     \centering
%         \includegraphics[width=1\linewidth]{chapters/images/20_biom_beta_recovery.png}
%     \label{fig:Synthetic_timeshifts}
%     \caption{Fitting synthetic timeshifts with the COMIND model. Solid lines are true trajectories/timeshifts, faint dotted lines are initial guesses, and dashed lines are final estimates.} 
% \end{figure}
\begin{figure}
\centering
\begin{minipage}{.5\textwidth}
  \centering
  \includegraphics[width=1\linewidth]{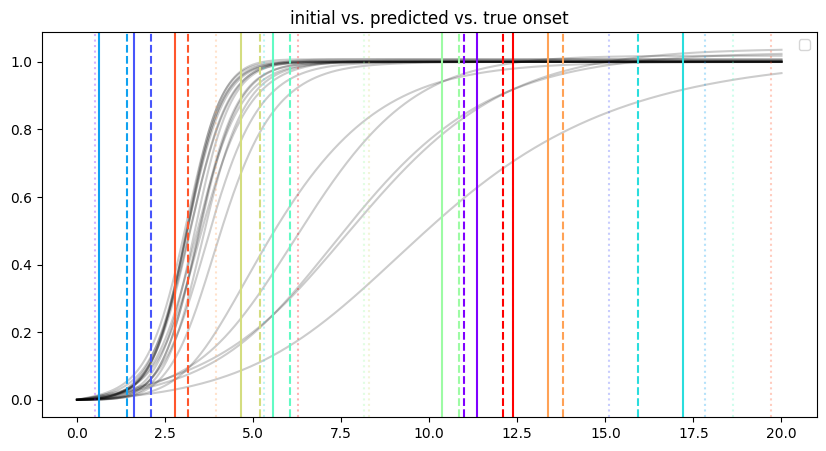}
  % \captionof{figure}{Fitting synthetic trajectories with the COMIND model. Solid lines are true trajectories, faint dotted lines are initial guesses, and dashed lines are final estimates.}
  % \label{fig:Synthetic_trajectories}
\end{minipage}%
\begin{minipage}{.5\textwidth}
  \centering
  \includegraphics[width=1\linewidth]{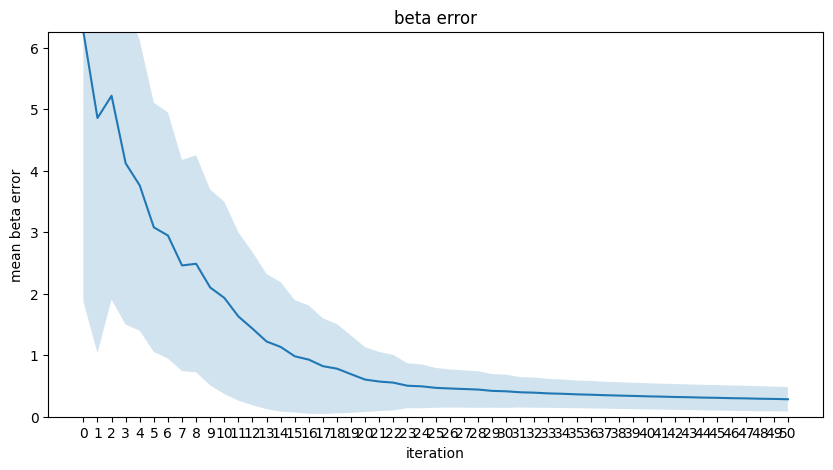}
  % \label{fig:Synthetic_timeshifts}
\end{minipage}
\caption{Fitting synthetic timeshifts (left) and evolution of timeshift prediction error (right) with the COMIND model. Solid lines are true trajectories/timeshifts, faint dotted lines are initial guesses, and dashed lines are final estimates.} 
\label{fig:Synthetic_timeshifts}
\end{figure}

% \begin{figure}
% \centering
% \begin{minipage}{.5\textwidth}
%   \centering
%   \includegraphics[width=1\linewidth]{chapters/images/20_biom_trajectory_fit.png}
%   % \captionof{figure}{Fitting synthetic trajectories with the COMIND model. Solid lines are true trajectories, faint dotted lines are initial guesses, and dashed lines are final estimates.}
%   \label{fig:Synthetic_trajectories}
% \end{minipage}%
% \begin{minipage}{.5\textwidth}
%   \centering
%   \includegraphics[width=1\linewidth]{chapters/images/20_biom_beta_recovery.png}
%   \label{fig:Synthetic_timeshifts}
% \end{minipage}
% \caption{Fitting synthetic trajectories (top) and timeshifts (bottom) with the COMIND model. Solid lines are true trajectories/timeshifts, faint dotted lines are initial guesses, and dashed lines are final estimates.} 
% \end{figure}

\subsection{Parkinson's Disease Imaging and Clinical Data}

The PPMI study employs a standardized high-resolution 3D T1-weighted volumetric sequence (MP-RAGE/IR-FSPGR) acquired on 3T scanners across ~50 international sites (sagittal 1.0x1.0x1.0 mm isotropic scan, $256^3$ FOV). We used 146 PPMI subjects with Parkinson's Disease who had at least one follow-up scan. T1-weighted scans were processed using FreeSurfer 7.0 longitudinal pipeline to harmonize cortical parcellations (Desikan-Killiany atlas, DK) across time points. We use cortical thickness measures in the 68 DK regions as our primary biomarker. There were a total of 504 observations, resulting in 34,272 unique imaging measures.

Additional measures included Montreal Cognitive Assessment (MoCA), Tremor-dominant sub-score (TD-Score) and Postural Instability-Gait Disorder Subscore (PIGD-score) from MDS-UPDRS-III symptom-specific items, Hoehn-Yahr (HY) stage and Neuronal alpha-Synuclein Disease integrated staging (NSD-ISS). HY and NSD-ISS progress in discrete steps from stage 0/1 to stage 6 (higher=more severe). Lower MoCA and higher TD/PIGD scores imply greater symptom severity.

\subsection{Connectome Model}

We constructed a canonical connectome from structural connectivity matrices of 794 subjects in the Human Connectome Project. These matrices were generated from preprocessed diffusion MRI data using MRtrix3 \cite{9-MRTrix}. Tractography was performed in an anatomically constrained manner based on T1-weighted MRI. We used constrained spherical deconvolution \cite{10-CSD} and intensity normalization to derive an initial set of ~40 million tractograhy streamlines (max length = 250, FA cutoff = 0.06). We applied SIFT2 \cite{11-SIFT} to reconstruct whole-brain streamlines 106. Streamlines were then mapped onto the 68 cortical Desikan-Killiany regions to produce subject-specific connectomes. These were then normalized and averaged. We excluded any self-connections from our connectome, setting the diagonal to zero for imaging-based biomarkers. 

\begin{figure}[h!]
\centering
\includegraphics[width=.9\textwidth]{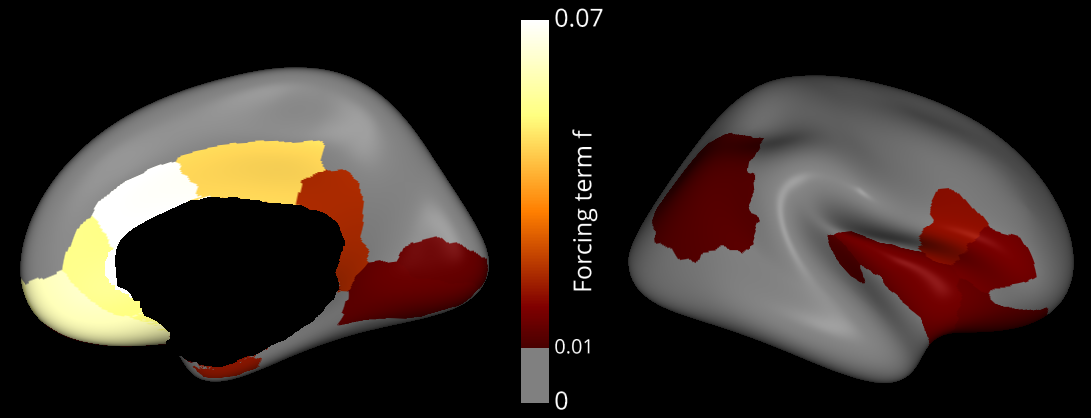}
\caption{Map of estimated external neurodegeneration components $\mathbf{f}$ in the Parkinson's Disease model.} 
\label{fig:Forcing_map}
\end{figure} 

\subsection{Synthetic Experiments}

We applied our inference algorithm for COMIND to 200 randomly generated samples with fixed hyperparameters $w_s, w_f$, and $w_c = \frac{\sigma_k}{\sigma_c}$, since $\sigma_k$ was constant across $k$ and $\sigma_c$ was known. We evaluated our performance based on trajectory recovery and sample timeshift estimation.

\subsection{Parkinson's Disease experiments}

We randomly selected 20\% of our subjects as the hold-out validation set. Next, we performed 3-fold grid search cross-validation to find optimal hyperparameters $w_s, w_f, w_c$. During the evaluation stage, we used the learned $\theta, a, \mathbf{b}$ to estimate $\beta$ on the inner evaluation set of samples. We selected hyperparameters leading to the smallest least-squares error between predicted trajectories and the evaluation set biomarkers. We used MoCA, TD-score and PIGD-score as the set of privileged biomarkers for conditioning patient timeshifts. To evaluate clinical utility, we estimated Kendall's Tau for ordinal correlation between the estimated disease time and (1) Hoehn-Yahr stage, (2) NDS-ISS stage. 
\begin{figure}[h!]
\centering
\includegraphics[width=0.85\textwidth]{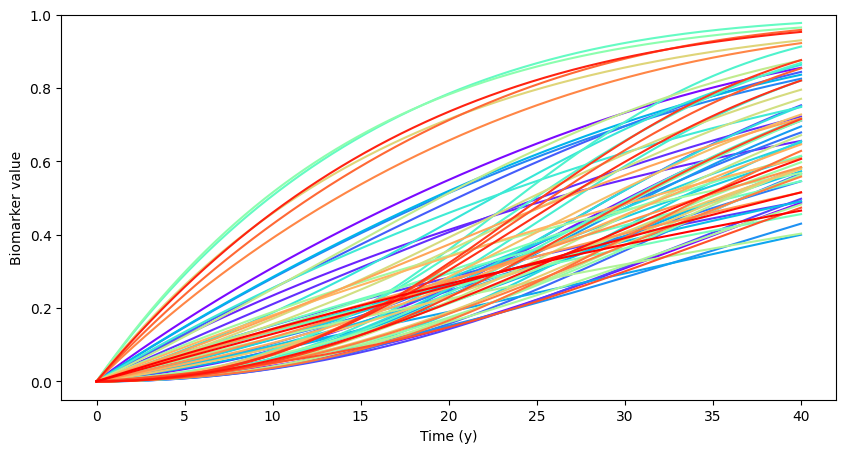}
\caption{The estimated normalized trajectories $\mathbf{x}(t)$ from the Desikan-Killiany regions' model of PD over the entire range of time considered.} 
\label{fig:All_68_trajectories}
\end{figure} 

\begin{figure}[h!]
\centering
\includegraphics[width=.85\textwidth]{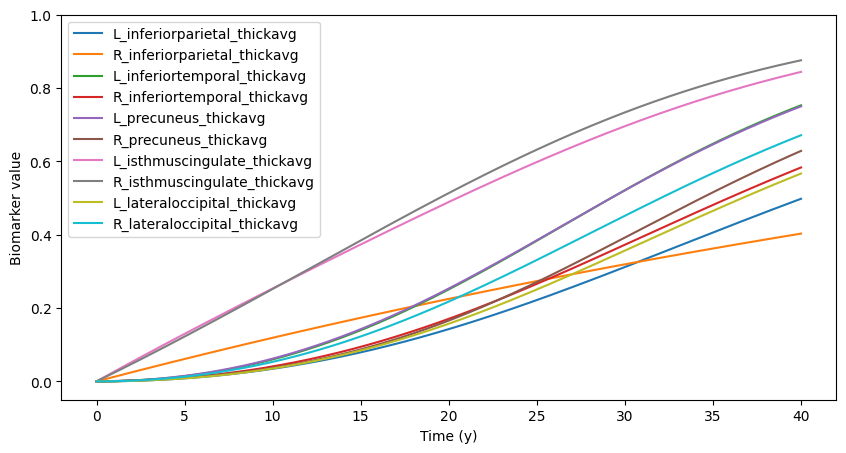}
\caption{Selection of trajectories from the 68-region PD model in $\textbf{Fig.\ref{fig:All_68_trajectories}}$ for regions showing greatest PD effect in \cite{12_ENIGMA_PD}} 
\label{fig:Big_model_10_trajectories}
\end{figure}

\section{Results}

\subsection{Synthetic results}
% \begin{figure}[h!]
%     \centering
%     \includegraphics[width=0.5\linewidth]{chapters/images/synthetic/20_biom_trajectories.png}
%     \caption{The biomarker trajectories based on predicted model parameters. Solid lines denote ground-truth trajectories while dashed lines indicated the fitted trajectory.}
%     \label{fig:enter-label}
% \end{figure}
% \begin{figure}[h!]
%     \centering
%     \includegraphics[width=0.5\linewidth]{chapters/images/synthetic/20_biom_timeshifts.png}
%     \caption{The predicted biomarker shifts \mathbf{\beta} from the model fit (dashed lines) compared against the true \mathbf{\beta} which the observed data was generated from.}
%     \label{fig:enter-label}
% \end{figure}

The COMIND model recovered trajectory parameters and sample timeshifts with reasonable accuracy, seen in $\textbf{Figs. \ref{fig:Synthetic_trajectories},\ref{fig:Synthetic_timeshifts}}$. Mean $\beta$ recovery error was 0.3 +/-0.2 on a set of timeshifts drawn uniformly from an imaginary 20-year range. Unsurprisingly, estimating disease time was easier for samples nearer in time to periods of relatively rapid biomarker decline, especially compared to samples at relatively later disease stages. 

\subsection{Parkinson's Disease Model}

\begin{figure}[h!]
\centering
\includegraphics[width=.9\textwidth]{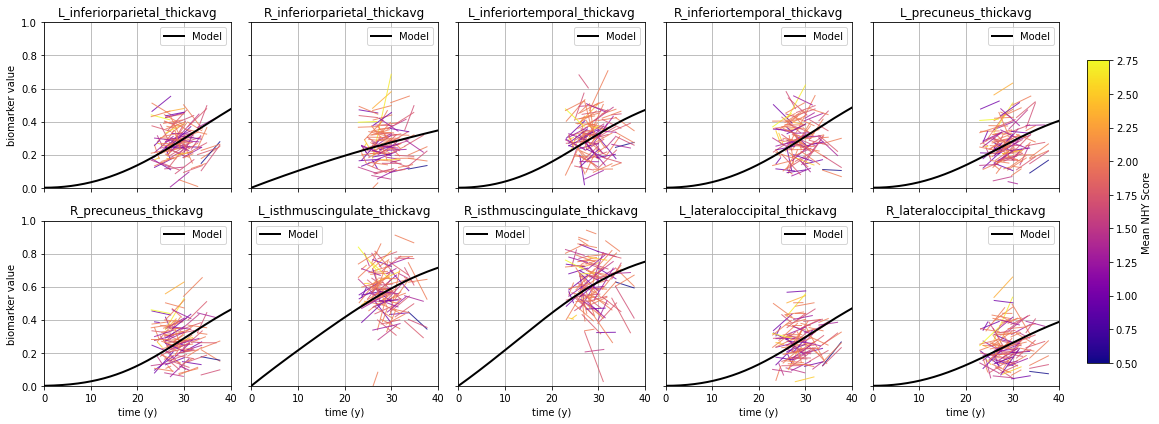}
\caption{Training subject trajectories overlaid on the model for 10 selected regions. Color-coding is based on the subject's Hoehn-Yahr stage.} 
\label{fig:Spaghetti_train}
\end{figure}

\begin{figure}[h!]
\centering
\includegraphics[width=.9\textwidth]{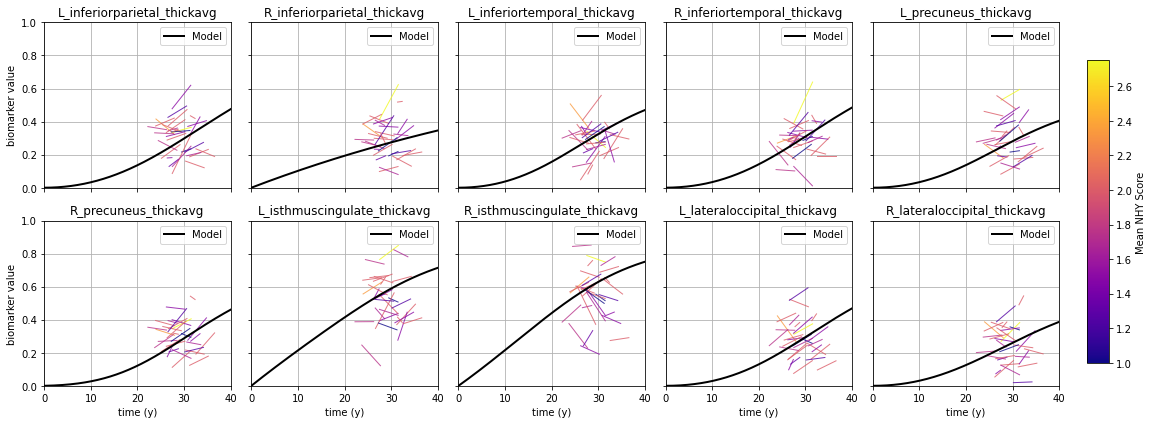}
\caption{Validation subject trajectories overlaid on the model for 10 selected regions. Color-coding is based on the subject's Hoehn-Yahr stage.} 
\label{fig:Spaghetti_val}
\end{figure}

\begin{figure}
\centering
\begin{minipage}{.5\textwidth}
  \centering
  \includegraphics[width=.85\linewidth]{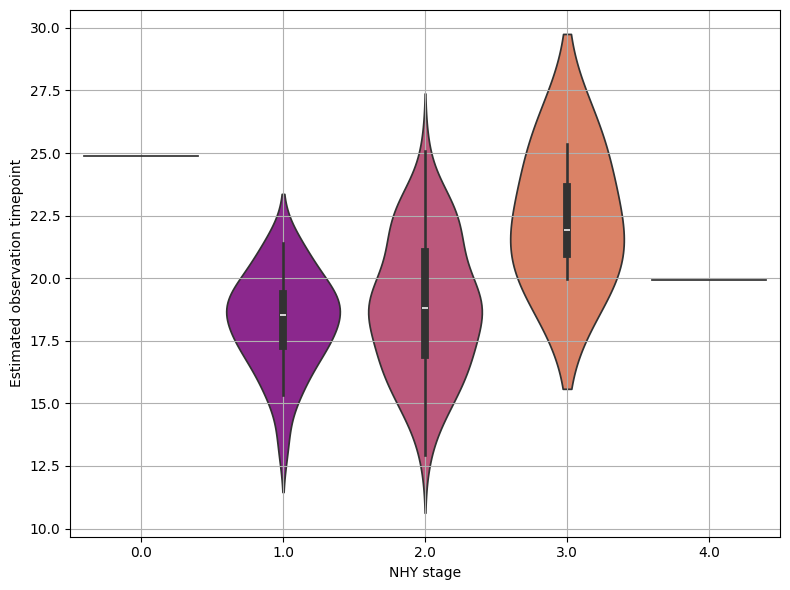}
  % \captionof{figure}{Distribution of estimated disease-time (timeshift) by Hoehn-Yahr stage.}
  % \label{fig:Violin_NHY}
\end{minipage}%
\begin{minipage}{.5\textwidth}
  \centering
  \includegraphics[width=0.85\linewidth]{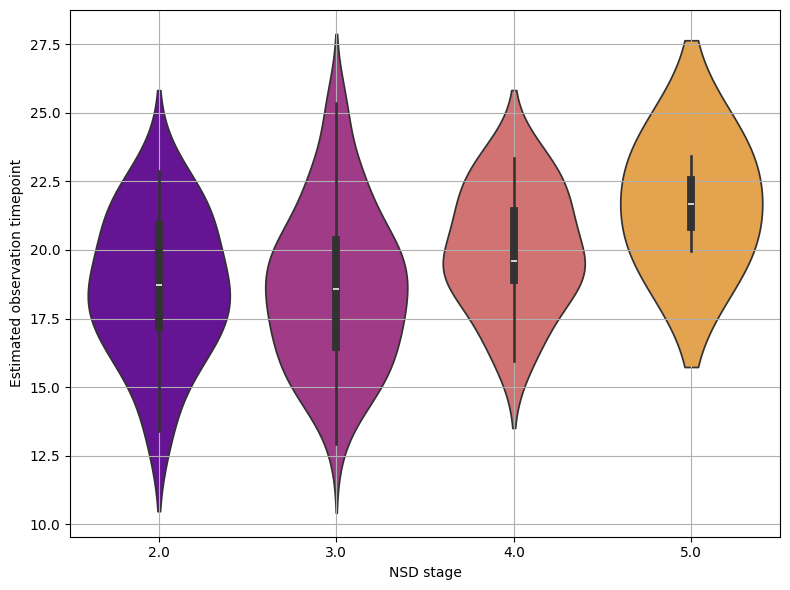}
  % \captionof{figure}{Distribution of estimated disease-time (timeshift) by NSD-ISS stage.}
  % \label{fig:Violin_NSD}
\end{minipage}
\caption{Distribution of estimated disease-time (timeshift) by Hoehn-Yahr stage (left) and NSD-ISS stage (right). Based on the validation cohort.} 
\label{fig:Violin_NSD_NHY}
\end{figure}

The Parkinson's Disease model yielded an external accumulation term $\mathbf{f}$ that was ~45\% sparse, and ~80\% sparse when setting all regions' entries below $0.15 \times \text{max}$ to zero. $\textbf{Fig.\ref{fig:Forcing_map}}$ illustrates this; interestingly, the remaining regions are loosely overlapping the standard salience network. Trajectories of the 68 Desikan-Killiany atlas and their subset corresponding to 10 regions most anatomically affected by PD \cite{12_ENIGMA_PD} are shown in $\textbf{Figs. \ref{fig:All_68_trajectories},\ref{fig:Big_model_10_trajectories}}$. Experimentally, the optimization approach of switching between NM and L-BFGS optimizers yielded monotonic decrease in least squares error (LSE) over 20 EM steps, with LSE reduction by a factor of 0.5. Goodness of trajectory fit can be seen in the "spaghetti plots" in $\textbf{Figs. \ref{fig:Spaghetti_train},\ref{fig:Spaghetti_val}}$.  Kendall's Tau for timeshift correlation was $\tau=0.10\; (p=0.19)$ for Hoehn-Yahr stage and $\tau=0.22\; (p=0.0053)$ for NSD-ISS stage. Distributions for each stage are displayed in $\textbf{Fig.\ref{fig:Violin_NSD_NHY}}$


\section{Conclusion}

We have presented COMIND: a progression model of neurodegeneration that combines diffusion-like and logistic accumulation dynamics with the network propagation hypothesis in brain disease. The method's parameter space is made deliberately low-dimensional to enable fitting on relatively small longitudinal MRI cohorts with potentially low disease-related signal. We demonstrate that the approach leads to stable and neurologically plausible results on anatomical MRI data from the PPMI cohort - a notoriously low-signal dataset in which researchers often struggle to find disease effects. It is especially promising that the NSD-ISS staging system appears to be in good agreement with our anatomically-predicted stage; the NSS-IDS is arguably the most comprehensive staging approach in Parkinson's Disease to date, incorporating genetic, DAT, and clinical information to identify patient stage. We also note that some extensions of the method are currently in progress, including the addition of subtyping via the clustering of $\mathbf{f}$ and adding dynamics to the connectome itself.

\section{Acknowledgments}

This work was supported by the Michael J Fox Foundation grant MJFF-021683, Multimodal Dynamic Modeling and Prediction of Parkinsonian Symptom Progression

% \printbibliography

\bibliography{LMID_bibliography}
\bibliographystyle{splncs03_unsrt}

\end{document}